\documentclass{article}
\usepackage[authoryear,round]{natbib}
\title{Self-consciousness and personal identity in quantum panprotopsychism}
 \author{Rodolfo Gambini
 \thanks{Instituto de F\'{\i}sica, Facultad de Ciencias, Igu\'a 4225, esq. Mataojo,
 11400 Montevideo, Uruguay.}
 \and
 Jorge Pullin\thanks{
 Department of Physics and Astronomy, Louisiana State University,
 Baton Rouge, LA 70803-4001, USA.
Corresponding author. Email: pullin@lsu.edu}}
\date{January 2025}
\begin{document}
\maketitle
\begin{abstract}
    In previous papers, we demonstrated that an ontology of quantum mechanics, described in terms of states and events with internal phenomenal aspects (a form of panprotopsychism), is well suited to explain consciousness. We showed that the combination problems of qualities, structures and subjects in panpsychism and panprotopsychism stem from implicit hypotheses based on classical physics regarding supervenience, which are not applicable at the quantum level. Within this view, consciousness arises in entangled quantum systems coupled to the neural network of the brain. In entangled systems, the properties of individual parts disappear, giving rise to an exponential number of emergent properties and states. Here, we analyze self-consciousness as the capacity to view oneself as a subject of experience.  The causal openness of quantum systems provides self-conscious beings the ability to make independent choices and decisions, reflecting a sense of self-governance and autonomy. In this context, the issue of personal identity takes a new form free from the problems of the simple view or the reductive approaches.    
\end{abstract}
\eject
\section{Introduction }

Panpsychism posits that fundamental physical entities possess mental-like qualities. This concept has ancient roots and has evolved over time.  In modern times, it was proposed by Spinoza and Leibnitz but was taken up with particular interest in the early twentieth century by William James, Alfred North Whitehead, and Bertrand Russell.
It was recently shown \citep{c1,c2} that a vision based on the scientific structuralism of van Fraassen, influenced by Hertz, Mach, Poincaré and others, together with the adoption of an ontology of events and states inspired in the axioms of quantum mechanics and implicit in most scientific theories, allows establishing a version of panpsychism free from the combination problem. The palette and grain combination problems of panpsychism and panprotopsychism arise from implicit hypotheses based on classical physics about supervenience. Such proposals are inappropriate at the quantum level, where an exponential number of emergent properties and states arise, leaving no trace of the properties of the parts of the system.  

Van Fraassen's interpretation of scientific structuralism considers physical descriptions as results of abstracting and idealizing an external reality, which we can only refer to in an indexical way, essentially pointing out the object in question. For him, starting from real objects, those that are correlates of observable phenomena that are not necessarily mathematical in their intrinsic nature, we attribute them mathematical properties through measurements. In the case of the planets, an example would be their location on the celestial sphere as seen from Earth. From the perspectives that the phenomena offer, adopting a heliocentric view, an empirical description is constructed, such as Kepler's laws, based on carefully collected astronomical data. The theoretical work that remains to be done is to obtain the empirical structure given in Kepler's laws from a model of the Solar System and the laws of dynamics. He describes the procedure as follows \citep{vfr2} p.6 ``The theory to phenomena relation displayed here is an embedding of one mathematical structure in another one. For the data model, which represents the appearances, it is a mathematical structure. So, there is a ‘matching’ of structures involved; but is a ‘matching’ of two mathematical structures, namely the theoretical model and the data model.”

All fundamental physical theories, whether classical like general relativity or quantum mechanical, share common elements. They consist of a mathematical framework, an interpretation, and rules of correspondence that link the mathematical formalism to observable phenomena. These theories offer mathematical descriptions of specific physical systems that start in a particular state and evolve to produce events that represent observable phenomena. We can associate certain mathematical objects with these states and events. The mathematical regularities derived from the description of the theory will hold true without exception, although in quantum theories they exhibit a probabilistic nature.

Events became central with the rise of quantum mechanics in the early twentieth century. This new theory expands our understanding of nature into a new realm of reality: the microscopic. It encompasses phenomena at the atomic and molecular levels, including those crucial to biological life, such as the quantum properties of large molecules called proteins. Since larger objects are made up of quantum entities such as atoms and molecules, quantum effects extend beyond the microscopic realm and are essential for explaining the physical properties of solids and liquids. Our current perspective is that classical physics is just one way to study certain macroscopic properties of quantum systems that have lost their coherence, a common occurrence in many systems at room temperature. Nevertheless, the world, when described from a third-person perspective, is fundamentally quantum in nature. The axioms of quantum mechanics refer to fundamental concepts such as systems, states, events, and their defining properties that correspond to observable phenomena. From this ontology, objects and events can be seen as the fundamental components of reality. In the quantum framework, objects are represented by systems in specific states. Events are the observable entities that manifest in measurements, while states represent the disposition to produce those events \citep{c1}.

In classical physics, current events are completely determined by the past. However, in our quantum world, this determination is only probabilistic. If we do not assume that the physical realm encompasses everything, then a probabilistic description allows for the possibility of new causal influences. For instance, phenomena from the realm of experience could play a crucial role in the natural world, as suggested by theories like panpsychism. Reducing consciousness to mere physics is neither necessary nor likely achievable. However, it is essential to understand the physical foundations of consciousness and the brain processes that enable it.

An ontology of events based on quantum mechanics could provide the following interpretation: Events in the external world are subject to a physical description, but are not necessarily exhausted by it. At least some events in our brain could be directly accessible as perceptions or qualia. The main difference would lie in the way we access events: first-person access for the mental, and access in third-person for the physical.  This suggests a form of panprotopsychism, where events and states would have intrinsic properties that, although not mental in the usual sense, when properly organized, give rise to consciousness in complex creatures such as us.

Bertrand Russell is both a scientific structuralist and a panprotopsychist. He argues that physics solely describes structures and causal connections, asserting that the matter in a given location consists entirely of the events occurring there. He further claims that this perspective on matter suggests that: “We no longer have to deal with what used to be mysterious in the causal theory of perception: [how] a series of light or sound waves... suddenly produces a mental event apparently totally different from them in character” \citep{Russellmatter}, p. 185. In Russell's view of neutral monism, the physical and the proto-psychic are simply two ways of approaching the same underlying entity: the event.

Here we would like to extend the analysis of previous papers, based on a form of quantum panprotopsychism that was applied to the different aspects of the combination problem of panpsychism: the combination of qualities, structures and subjects \citep{c1,c2}. We will analyze here the nature of self-consciousness and personhood, its meaning and identity over time. In section II we recall how quantum mechanics could solve the combination problem of panpsychism. In section III we describe the role and nature of subjects in quantum panprotopsychism. In section IV we analyze the three levels of subjectivity corresponding to arbitrary physical objects or proto subjects of experience, animal consciousness, and personal self-consciousness. In section V we discuss the issue of personal identity in this context. Finally in the section VI we conclude with some remarks and discussing possible future developments.

\section{Recalling how quantum mechanics can solve the combination problem of panpsychism}

\citet{strawson} p. 3 considers that the common belief that conscious experiences are basically inexplicable in physical terms results from the assumption that “The physical is, in and of itself, something fundamentally non-experiential in nature.” For a panprotopsychist, the intrinsic nature of the physical is experiential and therefore consciousness is naturally explainable. Understanding panpsychism through the lens of classical physics is problematic, as it fails to account for how complex minds emerge from interactions among numerous entities that might possess micro-consciousness. The challenge, known as the combination problem, as Seager \citep{seagersep} notes, lies in making sense of how "little" conscious subjects with their micro-experiences can combine to form a "big" conscious subject with its own experiences. This issue affects all aspects of our conscious life, relating to both our subjectivity and perceptions, and is viewed by panpsychists as a significant challenge to their perspective. 

The problem results from the implicit hypothesis that, as in classical physics, the properties of the parts remain unchanged in the whole and that the properties of the whole are nothing more than functions of the properties of its parts --- they are given by them ---.The classical physicalist program is to explain consciousness in terms of supervenience. Two classical systems are equal if their parts, and the relationships between them, are also equal. \citet{chalmers4} has analyzed the forms that the combination problem takes in the panpsychist approach, identifying three subproblems based on different aspects of phenomenal consciousness: its subjective character ---it always involves a subject---, its qualitative character ---it involves a great number of different qualities---, and its structural character ---it organizes in very complex structures---. They are respectively called the subject combination problem, the quality combination problem, and the structure combination problem. 

 The subject combination problem is especially important when examining self-consciousness and personal identity. It concerns how microscopic subjects of experience come together to form macroscopic ones like us. Within this context, we can identify several subproblems: we wish to understand what subjects of experience are, how they combine, and how to resolve the summing-subject combination problem. This involves explaining how micro subjects contribute to the emergence of a macro-subject.
If one adopts an ontology based on classical physics, one needs a form of constitutive panprotopsychism, where each property of the whole is determined by the properties of its individual parts. As Seager \citep{seagersep} observes, if one adopts this point of view, ``the fact that I am conscious consists primarily in the fact that certain particles in my brain are arranged or interacting in a certain way... This is the form of panpsychism that suffers most acutely from the combination problem.''

In contrast, non-constitutive panpsychism acknowledges both micro- and macro-experiences, but asserts that micro-experiences do not ground macro-experiences. This version of panpsychism cannot be sustained within a classical physics framework. However, as \citep{lulegues, healey, teller} point out, it naturally aligns with the ontology suggested by quantum physics.
In fact, in quantum mechanics there exists what \citet{healey} Sec.4 calls ``Physical Property Holism". It establishes that there are physical objects ``not all whose qualitative intrinsic physical properties and relations supervene on qualitative intrinsic physical properties and relations in the supervenience basis of their `basic physical parts'". The emergence of new properties in a quantum context --—where events and properties are fundamental--— highlights the ontological novelty and non-separability that characterize quantum phenomena, particularly in entangled states \citep{c1}. These behaviors of entangled systems allow wholes to possess properties that cannot be explained solely by the properties of their individual parts. For example, in a model of two spinning particles entangled in the singlet or triplet state, certain components of the total spin of the two-particle system have a property that does not follow from the parts, since the individual particles lack defined spin components. These kinds of properties may be crucial for the emergence of consciousness in composite systems with multiple components. In addition, the number of emergent properties, that do not stem from the characteristics of its parts, that a total system can exhibit, increases exponentially with the number of components. For example, a system of $n$ entangled spinning particles may be in different pure states that depend on $2^n$ independent coefficients, with different properties of the total system and different dispositions or causal powers to produce events in other systems \citep{lulegues,c1,Davies3}. 

The ''holistic" behavior of many nonseparable quantum systems is not an exception; rather, it is the norm in composite systems where components interact or have interacted in the past. For example, the electrons in a multielectron atom are entangled. This entanglement is a common characteristic: interactions usually lead to entangled states in multi-component systems. Much of the world we encounter in our daily lives appears separable because its quantum nature is obscured by the widespread phenomenon of environmental decoherence, which produces the classical behaviors we observe in macroscopic objects coupled with their environment.

The holistic and probabilistic properties of quantum systems are key for the solution of the combination problems of panpsychism and panprotopsychism. In fact, these problems arise because of implicit hypotheses about the classical supervenience of the properties of the top systems from the properties of their parts. 
For instance, the quality combination problem would involve understanding how micro-qualities combine to yield macro-qualities, such as the particular sound of a musical instrument.  If the world behaved classically, it would correspond to the problem of how phenomenal qualities like qualia would accompany certain physical events, such as neuron firings. 

The preceding quantum analysis of states and events allows the problem to be posed in much more promising terms. Indeed, the quantum emergence of states and properties does not allow us to consider that the properties or states of the whole are composed of properties or states of the parts as in classical physics. Instead, in an entangled quantum system, the parts lose their individuality and, as in the previous examples of spin systems, only the properties of the whole are defined. In other words, in a quantum system capable of maintaining its entanglement, composed of systems of molecules \citep{fisher} or photons coupled to the neural system, macro qualities would appear without leaving a trace of the micro qualities of the entities involved in its constitution.

It has been argued that states cannot be considered real because their definition is ambiguous. For instance, when considering non-local systems, such as particles in entangled states that occupy different positions in spacetime, it becomes problematic to define a state at a specific time, as this notion depends on the chosen Lorentz reference frame. However, if we define the state by its disposition to produce events, it can be rigorously demonstrated \citep{gapo} that this disposition is uniquely defined, and the state in the Heisenberg picture only changes when events occur. The disposition to produce events with certain probabilities, regardless of whether they are spatially separated in the relativistic sense, meaning that they are not causally connected, is an objective property of states \citep{Pusey3}. This emphasizes the objectivity of states when described in terms of dispositions, giving them the same ontological status as events, which represent the observable aspect of reality \citep{c1,c2}.

\section{Subjects in quantum panprotopsychism}

Strawson is probably the contemporary philosopher who has reflected the most on the character of subjectivity and the nature of the subjects of experience. For him, there is no mystery about what it means to be an experiencing subject because we know what it is to be one. A self-experiencing subject \citep{strawson} p.2 ``is necessarily to experience being an experience subject, and in that sense to know what it is to be a subject of experience”. That does not mean that only beings endowed with self-consciousness are subjects of experience. Indeed, his concept of the physical is not far from ours. As we mentioned above, he is a panpsychist. The mysterious thing that remains to be explained would be, in this context, the relationship between the neurophysiological description and our experiences, which is manifested in the combination problem and which in our opinion requires a quantum treatment for its understanding. As shown in \citep{c2} the treatment of subjectivity best adapted to the quantum context is that of Whitehead, since it may be applied to elementary physical processes involving the relationship between proto-subjects and objects. According to \citet{whiteheadaoi}, when extended to any physical system, the most basic phenomenal form of experience would have an emotional character. He says \citep{whiteheadaoi} p.176: ``the basic fact is the rise of an affective tone originating from things whose relevance is given... about the status of the provoker in the provoked occasion.” This type of subjective activity is referred to as a 'prehension.' 
An occasion of experience is always correlated with events, whereas perceptions are closely related with quantum systems in certain states.  They are dispositions to produce events. In Whiteheadian terms \citep{WhiteheadProcess} p.19: ''A prehension... is referent to an external world, and in this sense will be said to have a 'vector character'; it involves emotion, and purpose, and valuation, and causation'' in the same way that states have dispositions to produce events.  Whitehead is a panprotopsychist, he does not attribute consciousness to all entities in which prehensions or events occur. For him, the occurrence of events in physical systems is preceded by the prehension of a proto-subject. 

The proto manifestations of occasions of experience, that is, prehensions, in simple quantum systems extend from the preparation of a system in certain state through the measurement of the system. As it was discussed in \citep{c2}, the preparation allows to put the system in a particular state and is associated with the measurement of a previous system leading to an observable event. When a system in a state is measured, it leads to an event resulting from a random choice that ends up changing or annihilating the previous state.  Schematically, in a preparation, the events put the subject in a certain state; in a measurement, the subject in a certain state produces an event in the object at the end of a phenomenal process. In the former, a prehension is born with the event that characterizes its properties ---in particular its phenomenal qualities---, in the later it dies with the production of an event through its interaction with other physical systems. The first can be associated with a sensitive process, it produces a phenomenal quality, a proto perception; in the second, the phenomenal process culminates with an action, a proto volitional act corresponding to the choice of an event. Both processes were analyzed in detail in previous papers.

Quantum states become manifest through the production of events during measurements, where quantum indeterminism reveals itself. Each event is preceded by the prehension of a proto-subject of experience. A crucial aspect of this process ----its peak---- is the selection of one possible outcome during the measurement. In measuring fundamental physical properties, such as an electron's position, these choices are fundamentally random. Prehensions correspond to the phenomenal counterparts of the processes that precede measurements, and follow preparations.

The fundamental components of a conscious volitional act in higher conscious beings should resemble those previously discussed: an 'object' --—the brain--— and a subject’s prehension defined by the state of a specific set of neurons linked to a quantum system in an entangled state. For example, in Fisher's model \citep{fisher}, this might involve a multi-entangled set of Posner molecules, where a prehension occurs, activating an event: the firing of certain neurons that initiate an action. In short, a subject's prehension represents the phenomenal aspect of a system in a particular state, and its lifetime spans the interval between the event that prepares it and the event it ultimately produces in another system.

Two proto subjects can merge into one if they interact and become entangled; in this scenario, the individuality of the parts is lost, resulting in a new subject that corresponds to the system in an entangled state. On the other side, if, for instance, two spinning particles do not interact, they remain unentangled, allowing the classical ontology to apply. In this case, the final system will continue to consist of the two proto-subjects, each maintaining its individuality.

Once again, if we attempt to extrapolate from elementary measurements or preparations to a mental process, the quantum system will be represented by a multiparticle entangled state. Some components of this state would interact with the neural system, inducing or suffering a change in the state of the quantum system. If it is a preparation, it would manifest itself as a new perception. If it is a measurement, it would manifest itself as a volition.  As we have pointed out, when the total quantum system is entangled, the individual states of its components lose their distinctive character, making the only relevant subject the one corresponding to the entire entangled system. This point has been emphasized in the analysis of the quality combination problem. Once the total entangled system is identified as the subject of successive experiences, the subject summing problem can be resolved in the same manner as the other combination problem, thanks to quantum ontology \citep{c1,c2}.

\section{Proto subjectivity, consciousness and\\ self-consciousness}

As we have mentioned above, any physical system presents proto-subjectivity having mind-like properties like prehensions. Proto subjects, conscious beings, and self-conscious beings share some properties and develop others while evolving from the inanimate world to biological to personal self-conscious beings. \citet{strawson} has observed that a self-conscious being may be considered as having a set of properties that gradually appear during its development. He says ``I propose that the mental self is ordinarily conceived or experienced as (1) a thing or entity, in some robust sense (2) a mental entity, in some sense (3, 4) a single entity that is single both (3) synchronically considered and (4) diachronically considered (5) ontically distinct from all other things (6) a subject of experience, a conscious feeler and thinker (7) an agent (8) an entity that has a certain character or personality" \citep{strawson1997}, p. 19.

A proto subject is a system in certain state, which in previous papers we have called an object when considered from a third person perspective. It is always correlated with certain prehension with a perceptual or emotional tone, and in that sense is a subject of experience---a mental thing--- and finally, due to its disposition to produce events on other systems, is an agent. Therefore, proto subjects may be considered as things that are subjects of experience and have causal powers, having therefore certain individuality manifested in its dispositions to act. If the state cannot be put as a tensor product; that is, if it is not a factorable state \citep{Peres3}, it is singular. The state of a subject is supposed to have no independent portions. It is characterized by its disposition to produce events through arbitrary measurements and is uniquely defined. It satisfies Strawson’s condition of having the “kind of strong unity of internal causal connectedness that a single marble has as compared with the much weaker unity ...found in a pile of marbles” \citep{strawson1997} p.345. However, contrary to what happens with more evolved forms of subjectivity they are not necessarily ontically different from other proto subjects. They lack the uniqueness of higher forms of consciousness. For example, two identical spinning particles in the same state $|z,up>$ are indistinguishable. The only distinction that two electrons have results from their states, being impossible to decide if it is electron $1$ or $2$ that is in one of these states. In what refers to its persistence in time, a prehension persists from the preparation of the state until its measurements. After the measurement, the state changes and a new prehension is generated or is annihilated. The observable magnitudes in quantum mechanics are given by operators that act on the Hilbert space of the system. If the system is subject to a non-destructive measurement of certain magnitude $m_1$ corresponding to the operator $O_1$, the magnitudes $m_2 \ldots m_p$ corresponding to observables associated to the operators $O_2 \ldots O_p$ that commute with $O_1$ are preserved by the measurement. One can say that there is a partial preservation of the proto-subject properties and therefore the subject may partially persist beyond a measurement. Following \citet{whiteheadmot}, an individual physical object, such as a piece of quartz, could be considered to be characterized by the set of events produced by its interactions with the environment, but such an individual is composed of many disconnected subjects of experience . Its persistence does not imply the persistence of a single subject of experience.

\subsection{Consciousness}

In line with panpsychism, conscious intentionality gained new significance with the advent of life and the evolution of higher animals. Key innovations in animals include motility, perception, and emotion, with their openness to the world a central characteristic. In addition to the urgency shared by all living beings, which is related to their fundamental drive to preserve and enhance their lives, animal life introduces motility and "the interposition of distance between urgency and attainment" \citep{jonas}, p. 101. Identifying prey requires enhancing their perception, locomotion, and maintaining focus on the prey. In fact, in the last few years, the conviction has increased that consciousness is a widespread phenomenon in animal life \citep{low}. In the Cambridge Declaration on Consciousness it is established that: “The absence of a neocortex does not appear to preclude an organism from experiencing affective states. Convergent evidence indicates that non-human animals have the neuroanatomical, neurochemical, and neurophysiological substrates of conscious states along with the capacity to exhibit intentional behaviors. Consequently, the weight of evidence indicates that humans are not unique in possessing the neurological substrates that generate consciousness. Non-human animals, including all mammals and birds, and many other creatures, including octopuses, also possess these neurological substrates.”

Animals exhibit a direct, non-reflexive form of consciousness; they seem to possess what is termed as {\em direct inner acquaintance} \citep{metzinger} or {\em direct inner access} to their experiences \citep{persons}. The concept of acquaintance was introduced by Russell, who states that we have acquaintance with anything we are directly aware of, without the need for any inferential process. As  \citet{russell5} (p.108)  explains, ``I say that I am acquainted with an object when I have a direct cognitive relation with that object, I am directly aware of the object itself. When I speak of a cognitive relation here, I do not mean the sort of relation which constitutes judgment, but the sort which constitutes presentation.'' 
Lynne \citet{lrb} attributes a rudimentary first-person perspective to animals for their ability to intentionally and consciously interact with their environment. Its rudimentary perspective places it as an entity in an environment that it can perceive and with which it can interact. A newborn human also has a rudimentary first-person perspective and does not develop a sense of self until the middle of the second year.

Inner acquaintance is present in very young infants and many animals. They have consciousness but not self-consciousness. Wittgenstein in his {\em Tractatus} \citep{wittgenstein} uses the following metaphor: consciousness is for the infant ``like that of the eye and the field of sight. But you do not really see the eye. And from nothing in the field of sight can it be concluded that it is seen from an eye [italics in original]''. And as Choifer adds p.339: ``For the one who is at the origin [with direct inner acquaintance] there is no understanding of him/herself as a subject (of one’s own mental states), i.e., he/she is not yet a reflectively conscious subject. The one who is at the origin is not in the world but one with the world. The ‘origin’ perspective constitutes a unique, subjective first-hand encounter with a phenomenon. However, the price for this uniqueness is that one cannot, without stepping outside (i.e., switching to the ‘outside’ perspective) compare or share one’s unique and private point of view with anyone else’s. In this sense the point of view from {\em origio} is mute.'' It is a perspective without any conceptual content \citep{persons}. However, most animals possess not only sensitivity --—they are sentient--— but also the ability to perceive, meaning that they can discern whether the world is in a certain way or another. This capacity for perception is crucial to the lifestyles these animals adopt, enabling them to recognize potential prey or threats, for example. 

Many cognitive scientists have recently emphasized the significance of animal perceptual abilities. Butler summarizes some of these abilities as follows:``Among the recently revealed high-level cognitive abilities of various species of birds are a number of manifestations of working memory, including delayed-match-to-sample, episodic memory, transitive inference, and multistability. Additional avian abilities include category formation, language use and numerical comprehension, tool manufacture and cultural transmission of tool design, theory of mind, and a high level of Piagetian object permanence.” \citep{Butler} p.445. Various phenomena related to perception have been observed in animals. For example, the phenomenon of multistability allows ambiguous figures, such as the Necker cube, to be perceived in one of two (or more) configurations at any given moment, demonstrating that  intentional perceptual processes exhibit a certain universality beyond humans. However, while birds can make tools, they do not plan or teach. Only in humans ''teaching and learning have been generalized" \citep{Gazzaniga} p.29.

The formation of categories, which requires advanced cognitive skills, has been demonstrated in pigeons by \citet{watanabe}. The pigeons were initially trained to distinguish between pairs of abstract paintings by Picasso and Monet, with reinforcement provided for one set of Picasso works and another for Monet. They subsequently learned to differentiate novel pairs of Picasso and Monet paintings and responded similarly to Picasso paintings as they did to those by Braque and Matisse while linking Monet's works to those of Renoir. Similar clustering abilities have also been observed in other birds and mammals \citep{Butler} p. 446. In addition to these perceptual skills, pigeons exhibit intentionality, demonstrating agency and goal-directed behavior. As Lynne  \citet{lrb} observes, most animals meet the ``two conditions for an entity to have a rudimentary first-person perspective: (1) the entity is sentient, and (2) the entity has intentionality.”

There seem to be no distinct boundaries beyond which direct inner awareness and certain mind-like activities are absent. Nervous systems are found not only in vertebrates, but also in mollusks, arthropods, and even worms. For example, the Caenorhabditis elegans nematode has 302 neurons and approximately 7,000 neural connections. This worm has been extensively studied to investigate its neural functioning and its connections to very basic forms of consciousness. One such study focused on examining synaptic connections and changes in its olfactory capabilities \citep{heaven}.

In conclusion, varying degrees of sensitivity and goal-directed behavior can be observed in all living organisms. If we acknowledge that forms of interiority can exist even in primitive developmental stages, it follows that a comprehensive ontology of the physical world should include some form of phenomenal experience within non-living things.

As it was discussed in previous papers quantum panprotopsychism may solve the different aspects of the combination problem including the well-known quality, grain and subject-summing combination problems of panpsychism. 
Quantum indeterminism also helps to understand how the phenomenal and the physical \citep{c1,c2} could interact allowing the complex structural organization of perceptions in space and time. Indeed, the improvement in the ability to structure our perceptions would induce greater adaptive capacities and allow the action of natural selection. The explanation rests on the existence of systems in entangled states in which their parts lose identity and manifest themselves in an exponential number of possible quantum states of the complete system. The preparations and measurements of these states would give rise to subjects capable of perceiving and acting. This requires that the brain be capable of hosting an entangled quantum system coupled with the nervous system.

Returning to the Strawson enumeration, animals are  conscious beings and subjects of experience. If models like the one proposed by Fisher apply to realistic situations in the brain, their phenomenal properties ---their prehensions--- are correlated with systems in certain states. A succession of such elementary subjects of experience, of systems in certain states with certain properties and choices, constitutes a conscious individual. It is correlated with prehensions with perceptual and emotional tones, having the direct inner acquaintance and the intentionality mentioned above. It is a thing, as a piece of crystal or a cell are \citep{hospitable}. They are closed or open systems in a succession of states producing events in their interaction with other systems. They are things, that can undergo things and do things in Strawson's words \citep{strawson}. Their individuality results from identifiable forms of behavior preserved in time as their capacity to have unlimited associative learning \citep{gija} and memory. They are both physical and mental things. Their successive prehensions are related in the same way that the successive prehensions of proto-subjects going through non-destructive measurements are.  Strawson refers to a single entity that is single both synchronically and diachronically. When referring to the synchronic singularity, he is thinking of what he refers to as a hiatus-free mental property that in our terms would correspond to a prehension and its corresponding extension in time from preparation to measurement. Its uniqueness would result from the physical nature of the entangled quantum system that interacts with the brain. It is not clear for rudimentary conscious beings whether different subjects of experience at different moments would correspond to a single conscious being or to a succession of similar beings as happens with other things such as crystals or cells. However, they present psychological continuity since their past actions condition their psycho-physical development and are therefore manifested in their present behavior and they probably have an entangled quantum system coupled to their neural network that is preserved throughout their life. As we shall see, this diachronic uniqueness takes a stronger justification when self-consciousness is achieved.

In what concerns the fifth property in Strawson's enumeration that was clearly false for proto-subjects of experience like a particle in certain state, rudimentary conscious beings will be generally distinct from any other things. This includes members of the same species because of the dependence of each environment and the exponential richness of possibilities of the entangled states that provide the physical support of a given specimen. Any system in an entangled dispositional state is a subject of experience, including feelings and volitional behaviors, and therefore satisfies the sixth requirement. It is an agent that chooses how to respond to any situation showing intentionality when choosing randomly accordingly with the probabilistic limitations of quantum mechanics, among possible behaviors that start at decision acts in the brain that correspond to measurements of the quantum entangled system. Finally, rudimentary conscious beings like animals do not satisfy, as we shall discuss in what follows, the requirements for being persons and therefore having definite personalities beyond certain patterns of behaviors or moods.

\subsection{Self-consciousness}

Self-awareness involves the ability to pay attention to oneself and consciously know one's attitudes and dispositions. It requires awareness of the perceptions, emotions, and feelings, the recognition of our body and of its relationships with others. I am self-conscious if I am conscious that I am a conscious subject. It implies the recognition of our conscious contents. Not only the awareness of the contents but the certainty that they occur in me.
The one who acts and suffer through this body. Consciously observed perceptions are spatially structured and correspond to a particular perspective, centered on our body. The main property is that the subject recognizes itself as a distinct and constant point of reference with an identity as an individual separated from other individuals. In the recognition of our body a self intervenes. Although even fishes can distinguish their bodies, this does not mean that they recognize themselves as agents and subjects. Without conceptualization, that kind of consciousness does not seem possible. 

The distinction between their own body and other objects already exists in animals and results from their perceptions, as does their ability to recognize objects such as a stone or a ball by their effects and behaviors \citep{Gazzaniga}.  The informational structuring of perceptions probably has an evolutionary origin as it was discussed in \citep{c2} while the conceptual analysis have a linguistic one. Animals seem to have preconceptual abilities. For instance, they are able to correct their mistakes by means of perception, which implies in some cases the discrimination between what is correct and incorrect; and they do it despite their lack of language and a concept of truth or falsity. However, although animals may have the ability to classify objects based on the understanding of some relationships and analogies between them and consequently generate the same type of response towards them, for instance they could distinguish bananas and pineapples from a piece of wood, this does not imply that this animal has a concept of food \citep{dieguez}.

Neuroscientists have determined that self-consciousness is localized in a part of the left hemisphere of the brain closely related to the language centers. This has been verified, for example, with experiences of brain splitting where the corpus callosum that connects both brain hemispheres is cut and consequently the brain is composed by two disconnected parts. ''The patients all felt just fine after surgery and they themselves noticed no difference”. However, when light is flashed independently to the left and right hemispheres. if light is flashed to the left hemisphere the patient reports that he perceived a flash of light. When is flashed to the right hemisphere, the patient stated that he saw nothing, However if he could also respond by pressing a Morse code key he responded by pressing the key with his left hand. The left hemisphere contains the conceptualization center called the interpreter by  \citet{Gazzaniga}. In Husserlian terms, it provides a noema to each intentional perception. The interpreter is a module of the brain that explains events from the information it does receive. ''The interpreter module appears to be uniquely human and specialized to the left hemisphere. Its drive to generate hypotheses is the trigger for human beliefs, which, in turn, constrain our brain...Our subjective awareness arises out of our dominant left hemisphere's unrelenting quest to explain these bits and pieces that have popped into consciousness” \citep{Gazzaniga} p. 102. Therefore, empirical neurophysiological  analysis suggests that various phenomenal processes can coexist in the brain, of which we are only self-aware of some.

As noted by Daniel Dennett, ``Language gives our little brains a huge boost in cognitive power denied to all others" \citep{dennett2009} ) p.28. Primitive sound signals, such as alarm signals or announcements about food sources, existed among superior animals. Other human species, like the Neanderthals, also likely had primitive forms of language. With the development of language, humans transitioned from using alarm cries and summary descriptions to detailed descriptions of nature and the location of threats. This advancement allowed humans to develop superior organizational capabilities to respond to environmental opportunities and challenges. The development of language was accompanied by other capabilities, such as artistic ones, which were also present in rudimentary form in other humans like Neanderthals. These abilities resulted from a growing capacity for abstraction and interpretation. The pictorial images left by cavemen are evidence of the unique human ability to symbolize aspects of perceived reality through images. When we see one of these images in a cave or on a rock, we know that it was a product of human work.

The pictorial representation shows that its author has achieved a new way of relating with objects through the mediation of an idea. The resemblance between the object and the image is not exact. The image highlights only the aspects of the object that are significant enough to be generally identifiable by sight. This abstraction in the image manifests itself in two ways: simplification, which can be done in various ways by emphasizing different characteristics of the objects, and generalization, where the image does not depict a specific bison, but any bison \citep{persons}. As noted by Hans Jonas, this process involves two factors: perceiving similarities and the deeper task of conceptualizing our perceptions: “The image becomes detached from the object, that is, the presence of the eidos is made independent of that of the thing... [ a second detachment] takes place when appearance is comprehended qua appearance, distinguished from reality, and, with its presence freely commanded, is interposed between the self and the real whose presence is beyond command'' \citep{jonas} p.170. Here Jonas almost inadvertently shows how the self is identified through this abstraction process that involves elaboration and choice. The process of image creation implicitly implies the freedom of the artist to choose the elements that she will use in her representation. This is not unique and the approximation to the model can be achieved in many ways. The one that is finally captured in the image is a product of a choice. Conceptualization and freedom to choose therefore seem to play a fundamental role in the constitution of the self. To the generality of the image corresponds the generality of the name that through the phonetic sign is related with many individuals \citep{jonas}. Language and the capacity for symbolic representation arise accompanying the development and autonomy of the self.
   
The artistic and linguistic development is accompanied by a growing concept of ourselves. Children acquire the idea of ''me" through their social and linguistic interactions during their early years, paralleling the development of their abstraction capabilities. Learning a language and developing a first-person perspective of the world significantly expand our capabilities. As Dennett observes, placing our brains on a continuum with insect nervous systems at the low end, and dogs, dolphins, and chimps near us at the high complex end, overlooks the obvious fact that we are the only species that asks questions. This is directly related with the constitution of the interpretational modulus of the brain mentioned above which is typical of humans and is not observed in other animals. The supposed continuity of evolutionary processes does not deny the remarkable qualitative changes that accompany the development of knowledge and human culture. 

Every deliberate act has some purpose and assumes a self that has a past and a future. We probably do not know how to demonstrate the uniqueness and persistence of my “I”, but we certainly cannot act without assuming it. I am concerned with the self that I was and the self that I will be. This consciousness occurs in this body, a body from which I suffer, perceive, and act. Many representations, whether through an image or word, involve a search for the truth. In that sense understanding the origin of conceptualization is related to the double movement of seeking the truth and avoiding error that is essential for the preservation of the self. 
Recognizing that a certain judgment or act has been wrong and has led me to suffer its consequences leads to its rejection. That is, to the denial of what is false and the affirmation of what is true because of a deliberate act and not a mere product of a reaction and therefore to recognize error in its most basic form as a false perception. Perhaps the function of the image, like that of many primitive words, is the search to avoid false perceptions. As \citet{jonas} suggests, perhaps the image seeks to avoid confusion between similar species: let's say to distinguish between buffaloes and antelopes for which ``forms are compared, in this case of general images rather than of concrete individuals” p.181 for which one chooses to highlight some distinctive characters.

When we exercise our ability to observe our conscious states, we also observe that we have that capacity, we can analyze them phenomenologically and act on them. Discovering the self is not knowing it; for this, we need to see it in action.
The concept of self is dynamic, we observe it in action in the subject as possessor of thoughts, feelings, beliefs, experiences, and perceptions, of acting and interacting. In that sense, it seems difficult to talk about the self in an instant. Experiencing it is the sense of being a unique person with an inner life and continuity over time. With the ability to make independent choices and decisions, reflecting a sense of self-governance and autonomy.
As a subject of experiences, the self perceives, acts, and expresses itself conceptually and symbolically, based on its capacity for autonomous action. Develops values and beliefs, thoughts and behaviors, choices and decisions. The self is the subject of its intentional states; it can reflect on them and act freely and responsibly.  The subject who evaluates his goals and beliefs, or wonders about his future and his goals, choosing a course of action or a particular way of thinking.

Only persons are aware that they are subjects of experience. It is one thing to experience needs or fears, to pursue prey or flee, and another to have a first-person perspective. A first-person perspective only exists in a person, it is its essential characteristic. It requires conceptual elaboration. A concept of oneself as a pole of action and experience. To recognize its body, the animal uses its perceptive capacity; to express my needs or perceptions or desires, I need conceptual capacity. I can use the pronoun I to refer to the entity capable of perceiving itself as the subject of perceptions decisions and volitions. Not only about being a subject, but about feeling like a subject. It is the first-person perspective that allows us to be subjects conscious of their intentional states. Only people have this perspective \citep{lrb}. 

We choose to adopt at this point the "constitution view" which holds that self-consciousness, the ability to perceive oneself as a subject, is the distinctive property of persons, regardless of whether they are human or of another origin. For the constitution view “A self-conscious being is a fundamentally different kind of thing from anything else in the world.” p.31 of \citep{baker2}. Surely, such a capacity is not just a form of mental functioning. It must have a physical counterpart, which in the quantum models we have considered would probably require a mirror neural system capable of mapping the activity of certain regions of the brain and an entangled quantum system that interacts with the original neural system and the mirror one \citep{rizzolattietal}. 

Self-awareness is not only the ability to conceive of ourselves as agents endowed with intentionality and sensitivity. With it, a new form of relationship with the physical world appears, not only given by perceptions but in the interposition of the concept or image between the mentally manipulable products of our previous dealings with objects and the directly perceptible. Non-human animals have only biologically given goals that they cannot modify, human persons may evaluate and modify their goals. This strongly suggests that self-consciousness, self-determination, and free will are closely intertwined in persons.  If Samuel Alexander's  maxim "To be real is to have causal powers" \citep{alexander} is true, there should be something real ---an agent--- at the origin of the choices that accompany our reflective processes. Their nature would ultimately be determined by their free actions. The type of identity that persists will not be found in the self that we are at each moment. It will be found in the succession of acts that respond to the call of our objectives, existential concerns and moral responsibilities. ``Supremely concerned with what he is, how he lives, what he makes out of himself, and viewing himself from the distance of his wishes, aspirations, and approvals, man and man alone is open to despair'' \citep{jonas}. Thus, one cannot reduce the person to an isolated act of prehension or a succession of them. Persons are basically constituted in time in terms of conceptual elaborations and successive choices to act. 
 
The deterministic view of classical physics basically implies that our actions are an inescapable consequence of physical laws and our past, so it seems to prevent human agents from having the freedom to do otherwise. It also seems to prevent them from being the sources of their actions, and therefore it does not appear to leave a margin for us to be responsible of our acts. Determinism threatens this picture of freedom, because it seems to imply that there is only one possible path into the future, not many. ``Determinism therefore seems to prevent human agents from having the freedom to do otherwise, and it also seems to prevent them from being the sources of their actions. If either of these is true, then it’s doubtful that human agents are free or responsible for their actions in any meaningful sense'' \citep{mckenna}. 

As it was analyzed in \citep{c2}, the richness in the structuring of our perceptions requires some form of phenomenal efficacy leading to adaptive changes. This would allow the genetic development of better information processing capabilities through improved neural organization. The phenomenal qualities of consciousness must therefore have a certain degree of efficacy. This can be understood in physical terms in models where quantum mechanics plays a role and the prehesions associated with conscious states determine dispositions to action. Many neuroscientists question the effectiveness of consciousness based on observations such as the following: It has been observed on many occasions that movement precedes consciousness ``the (slow cortical) readiness potential (RP), originally termed Bereitschafts potential, is observed over motor regions in the electroencephalogram (EEG) as slow-building negative activity preceding movement onset... The so-called classical interpretation of the pre-movement activity observed in the RP is that gradual increases in negative brain potential over motor regions reflect a specific and goal-directed process for preparing the initiation of upcoming voluntary acts...” \citep{armstrong} p.1. This observation, together with classic physicalism and therefore a deterministic prejudice, has led them to conclude that the idea that our actions are free is a mere product of an a posteriori interpretation of the self. The statement is far from obvious: suppose that while I am writing this work a family member calls me to repair some domestic appliance, I will surely not answer the call with an automatic movement. For example, I can decide to finish presenting the idea I am developing before doing so. At some point I feel satisfied with that task, and I think I can get up. The reaction of movement is not necessarily immediate, the observation that I am able to attend to the other problem releases the automatic mechanisms that initiate the movement including the precise moment in which I begin to do so. The difficulty with the neurophysiological identification of volitional acts extends to all their manifestations not just movements. This conditions and restricts the ability to identify intentional acts. Introspective access is limited by the distortions that this process introduces in volitional activity. The difficulties of direct access to them are not surprising, given the absence of a physical counterpart to quantum leaps and the process of selecting an option. They are recognizable only through prior deliberations and subsequent conceptual, cultural or simply motor manifestations. Something similar happens with the type of experiences usually reported by neurophysiologists. The experimenter tells the subject: whenever you want, raise your arm. The subject, following what was requested, waits for a period of time until he considers that a sufficient period of time has elapsed and stops waiting, the brain then spontaneously initiates the requested movement. It would be more interesting to observe other types of actions, such as that of a person with little experience in cooking preparing a dish. The mere attempt to observe a decision-making under laboratory conditions at the request of an observer distorts the process, which does not result from a conscious decision associated with an act that has a purpose or a rational, ethical or aesthetic meaning. Recall that quantum systems always present some kind of top-down causation. As we have mentioned above this kind of causation is always present in systems in entangled states where the behaviors of the parts under measurements present correlations that do not result from the state of the corresponding part but of the whole \citep{GaPuEO, lulegues, healey,teller}.

Persons may evaluate and modify their goals. To be free is not to act randomly, but to choose between different lines of conduct attending different impulses or orientations, usually pre-established culturally in the form of beliefs, values or ideals. One chooses any time one thinks, composes, paints, helps someone else, also when we decide to stay without doing anything... we can always act in a different way. We always have the challenge of doing it better. The self feels dissatisfied in the face of a given set of answers and seeks until it finds others that quiet its dissatisfaction. Any creation is a choice \citep{persons}. Paraphrasing Foucault, one would say that not only is ``freedom the ontological condition of ethics" but also of aesthetics and of any form of symbolic activity \citep{foucault}.

If quantum indeterminacy plays a role in our mental processes, then personal freedom could imply more than mere random choices \citep{persons}. However, this would require identifying a kind of ethical and aesthetic orientation accessible to our consciousness. Although we do not attempt to solve this issue here, we note that we are capable not only of conceiving a range of potential aesthetic or moral choices, but also of evaluating their suitability for specific situations with a reasonable degree of accuracy. Often, we make choices that align with these evaluations, suggesting that this ability may be rooted in intuition, resembling perception more than rational analysis \citep{c1}. In a similar vein, Christine Korsgaard \citep{korsgaard} argues that (p.19) ``The capacity for self-conscious reflection about our own actions confers on us a kind of authority over ourselves, and it is this authority that gives normativity to moral claims.” 

Let us return once again to the properties that according to Strawson characterize a personal self. “The mental self is ordinarily conceived or experienced as (1) a thing or entity, in some robust sense (2) a mental entity, in some sense (3, 4) a single entity that is single both (3) synchronically considered and (4) diachronically considered (5) ontically distinct from all other things (6) a subject of experience, a conscious feeler and thinker (7) an agent (8) an entity that has a certain character or personality" \citep{strawson}, p. 19. It is necessary to emphasize here that the above list refers to the phenomenological nature of the self and that our aim is to relate it to the above metaphysical considerations about the nature of self-conscious subjects of experience. The list follows a first discovery of the self, made in our childhood when we observed that our thoughts are not accessible to others and that one is alone in one’s body. First, the self is something that has causal powers and undergoes things and does things. We do things and things happen to us. Second, our notion about the self results from our mental experience. Our certainty about its existence is based on our mental experience of the self. Without denying or affirming that it also has a non-mental nature. We always experience a single thing, and we never experience a plurality of selves present in a prehension. Fourth, I experience myself as a single thing persisting through time, through my different prehensions. However, for this last property, memory seems to play a crucial role in providing evidence of such a uniqueness. Fifth, it is ontically distinct in the sense that it is not like any other thing, and it is the characteristic thing that defines a person no matter what biological or physical character it has, human or extraterrestrial. The self constitutes persons \citep{lrb}. Sixth, it is a subject of experience, and it experiences that it is one. This is the starting point for the analysis of the self. Seventh, it is an agent; it defines itself through its choices and actions. Eighth, it is an entity that has a certain character or personality although it is not clear what part of that character is a property of the self and results from its actions and what part depends on biology, education, and life history.

Strawson, in his phenomenological characterization, distinguishes between two aspects of the self: the "I" as the subject of experience and the ''Me" as the objective aspect constituted by our self-image. This distinction, suggested by William James, differentiates between the self as the actor or knower and the self as "an empirical aggregate of objectively known things" \citep{persons}.
James' concept of the ''Me" is associated with the self-image we construct or wish to project. He posits that an individual can create different ``Me-selves" based on varying social roles. Many scholars, such as Tagini \citep{tara}, suggest that the 'I' relates to self-referential awareness tied to phenomenal consciousness rather than implicit representations. In contrast, ``Me"-related self-awareness is linked to the reflective mind arising from conceptual construction. For James, the "I" is that which at any given moment is conscious, whereas the ''Me" is only one of the things which he is conscious of. The "I" is the subject of prehensions and actions, while the "Me" is the explicit sense of self with unique, identifiable features forming one’s self-image or self-concept \citep{tara}. Despite being a conceptual construction, the ''Me" is special as it often refers, indirectly through others' judgments, to the various manifestations of our ''I" through behaviors. Thus, the ''Me-self" is a cognitive construction, essentially a ''theory" of the self. A constant challenge to our "objective" self-concept is maintaining a coherent and unified self-image \citep{epstein,persons}. Although the "Me-self" is a theoretical construct and is subject to various forms of error, it partially describes in first person properties of the organization of our brain that resulted from perceptions and acts of the ''I".

\section{Personal identity}

Eric Olson \citet{olson} lists a number of possible answers to the issue of personal ontology. ``What properties of metaphysical importance do we human people have, in addition to the mental properties that make us people? What, for instance, are we made of? Are we made entirely of matter, as stones are, or are we partly or wholly immaterial?... Are we substances ---metaphysically independent beings--- or is each of us a state or an activity of something else?'' The question of personal identity involves understanding what it means to be the same person from one day to the next. From the moment we discover our self, we become concerned with our permanence and the passage of time. My personal identity in this sense consists of the properties that I take to define me as a person. As we have discussed above, being a person requires having a first-person perspective, It requires conceptual elaboration. A concept of oneself. To recognize its body, the animal uses its perceptive capacity; to express my needs, perceptions, or desires, I need conceptual capacities. I can use the pronoun ''I" to refer to the entity capable of perceiving itself as the subject of perceptions or volitions. Not only about being a subject, but about feeling like a subject. It is the first-person perspective that allows us to be subjects conscious of their intentional states. Only people have this perspective. According to  \citet{korfmacher} solving the metaphysical problem of personal identity essentially requires answering how the phenomenon  by which ``entities like us” persist through time can be specified. We are focused on identifying the truth-makers of personal identity statements: what makes it true that an entity $X$ at time $t$ and an entity $Y$ at time $t'$ are identical, if $X$ and $Y$ are entities like us? 

\subsection{Personal identity: the characterization question}

Our "Me" concept evolves over time, along with our knowledge and beliefs. A similar evolution occurs with our "I" concept. Even when I consider myself a subject of experience and an agent. My brain is not the same and I am not the same agent that I was as a child or in my 20s. My knowledge and beliefs have evolved, as have my doubts and fears. What leads us to consider entities like persons at different times as a single persistent object? If one considers that persons at certain time are essentially characterized by their prehensions and choices at this time, one can recognize three levels involved in the process of perception and reflection eventually followed by an act of the “I” that provides a notion of personal identity consistent with the form of quantum panpsychism proposed here. The elements that characterize a self-conscious person at certain time are:

i) The information stored in the neurological organization of a brain capable of exchanging information with the sense organs and interacting with a body.

ii) The prehensions at this moment.

iii) When prehensions are followed by volitional or conceptual choices: the agent who chooses between the different possibilities posed by the topic in question and, with this choice, contributes to shaping the concepts and beliefs encoded in the brain and through its body influences other bodies.

Note that in the context of quantum panprotopsychism, for example, the one based on Fisher's model of cognition, each act would correspond to an interaction between Posner's entangled quantum systems of molecules and a set of neuronal synapses that would manifest itself in neural discharge events. Consequently, the acts would have a perfectly defined physical counterpart. However, due to quantum indeterminism, two identical brains after certain prehension could present different behaviors depending of the particular choice that may result from a quantum system in certain dispositional state.

The above mentioned levels determine someone’s personal identity at a certain time, and answers the so called characterization question \citep{schechtman}. This characterization is not far from the Lockean one that established that a person is “a thinking intelligent being, that has reason and reflection, and can consider itself as itself, the same thinking thing, in different times and places” \citep{locke} p. 335. It is about these three items that the question about its identity in time should be raised. As we have discussed previously, both animals and humans are conscious, but only humans become selves. This particularity of humans is related to their conceptual capacity and their ability to use highly structured symbolic systems such as languages. As suggested by Pickering \citep{pickering} (p.9) the self ``participates in the continual creative advance that, in bringing the past into the present, opens up a space of possibilities from which choice makes a bridge to the future.'' The central role of the self lies in these choices, but these choices do not occur in a vacuum. They are conditioned and informed by the contents of the brain (our memories, knowledge, and beliefs) and by our perceptions and the space of possibilities and dispositions present in our phenomenal experiences. It is around this self, responsible for its actions, where the different levels converge, that the question of personal identity over time must be raised \citep{c2}.

For Whitehead “The soul is nothing else that the succession of my occasions of experience, extending from birth to present moment” \citep{whiteheadmot} p.138. One can consider that the succession of occasions of experience have physical counterparts in states and events. Implicit in this definition is the agent behind the acts of choice allowed by the probabilistic quantum axioms. The states and events are the physical counterpart of selves, but if freedom does not reduce to random action, the self is not exhausted by them. Its most fundamental nature is being the entity able to reflect about itself and choose its own way of being.

The interiority of the subject becomes significant when the nomological deductive description of physics is insufficient to fully predict its behavior. It is this aspect of the self, as an agent of free choices compatible with the quantum probabilistic character of the subjects of experience, that places it in a unique sphere as the only entity whose transcendence of physical laws appears to be necessary. Freedom is the central manifestation of the self. In each action, the self has the option to exercise its freedom or not, but this option is limited by various forms of conditioning. This conditioning, along with our previous choices, ultimately determines our mental states and our willingness to act in a certain way. Such conditioning, whether derived from our biological nature, such as sexual drives and survival needs, or from social life, do not prevent free acts but limit them determining our dispositional state. When the disposition to a given behavior is strong, the "freedom to do otherwise" always present has a slim margin of maneuver \citep{persons}.  As noted by Korsgaard (p.96), the freedom involved in the acts of choice cannot be denied: ``The freedom discovered in reflection is not a theoretical property which can also be seen by scientists considering the agent's deliberations third-personally and from the outside. It is from within the deliberative perspective that we see our desires as providing suggestions which we may take or leave. You will say that this means that our freedom is not 'real' only if you have defined the 'real' as what can be identified by scientists looking at things third-personally and from outside [the same happens with reasons]. We need them because our reflective nature gives us a choice about what to do. We may need to appeal to the existence of reasons in the course of an explanation of why human beings experience choice in the way that we do, and in particular, of why it seems to us that there are reasons...reasons exist because we need them, and we need them because of the structure of reflective consciousness.” In particular, for these types of choice, the Aristotelian  phronesis is required \citep{aristotlenico}. It encompasses not only the ability to decide how to achieve a particular end, but also to reflect on and determine good ends that align with the overall goal of living well. 

The probabilistic nature of quantum mechanics makes this view compatible with a third-person physical description of the world.
It is crucial to understand that randomness does not inherently signify a lack of meaning. Information or meaning can still be conveyed through a sequence of symbols following a specific probability distribution. For instance, in the English language, letters occur with certain known probabilities, yet this does not hinder our ability to communicate effectively. Similarly, a probabilistic framework such as quantum mechanics does not preclude the presence of phenomenal causal properties underlying a sequence of events or measurement outcomes.

The quantum panprotopsychic view considers that the world is essentially phenomenal, but while prehensions in the subjects of experience have a physical counterpart in states and events, the selves as the agents that choose different options resulting from their reflection process allowed by the physical description, are not ruled by any physical law beyond the limitations imposed by the quantum probabilistic description of the outcomes. Beyond its conditioning, the self can develop through the exercise of its freedom and from its direct access to its states "from the interior" without requiring any measurement. Its actions are creative, producing objects for its use, inventing tools, symbols, and representations, and thus creating its own cultural and social world, distinct from nature. Because of the causal openness of quantum mechanics, having the complete physical information of a quantum system like us and of the environment with which we interact does not determine our future behavior, which also depends on our future quantum choices. The kind of choices we make are part of our personality. 

It is the subject capable of perceiving and evaluating its intentional states, reflecting on them, and acting responsibly. This is the subject of our choices, the one that evaluates objectives and contemplates the future. It is the source of action that abstracts, analyzes, and acts. Without self-consciousness, there would be no culture, and questions or statements about existence, purposes, or actions would be meaningless. Language would be rudimentary or nonexistent, and morality would be impossible, since self-awareness is necessary to evaluate the correctness of actions. Self-consciousness allows us to assess and alter our individual and social behavior and reflect on life, meaning, and destiny. 

{\em The question of the persistence of the self is not only related to the sequence of prehensions supported by the classical and quantum brain activity, but also includes the persistence of the agent responsible for free acts, and consequently, one should define persistence in terms of these elements: brains, quantum states and the events that prepare them, and the acts leading to the events involved in decision making, conceptualization, and symbolic expression.} The three levels that intervene in the process of reflection followed by a free action of the “I” are involved in the self’s manifestations, but the acts are the main cause of the development of our brain capacities and of its present activities, and therefore the defining elements of the self. As Cassirer observes in his study of the development of the self in mythical thought "It is not mere meditation but action which constitutes the center from which man undertakes the spiritual organization of reality. It is here that a separation begins to take place between the spheres of the objective and subjective, between the world of the I and the world of things. The farther the consciousness of action progresses, the more sharply this division is expressed, the more clearly the limits between I and not-I are drawn. Accordingly, the world of mythical ideas, precisely in its first and most immediate forms, appears closely bound up with the world of efficacy." 

\subsection{Personal identity: requirements and sources of evidence}

We therefore propose the following requirements for personal identity based on the above-mentioned characterization of a person:

i)	{\em Classical organizational connection (COC) } A being is organizationally connected at $t' >t$, with you as you are at $t$ just if and only if the neurological organization of its brain at $t'$ is causally connected with the neurological organization at $t$. That is, if both correspond to two stages of development of the same neurological organization.

ii) {\em Phenomenal continuity (PC)} There is phenomenal continuity between the subject $S$ at $t$ and $S'$ at $t'$ if and only if the prehensions of $S$ at $ t$ were the prehensions of $S'$. 

iii) {\em Identity of agents(IA)}  The agent $A$ at time $t$ is identical to $A'$ at $t'>t$ if and only if the acts of $A$ at $ t$ were acts of $A'$.

This definition seems to include purely phenomenal elements like prehensions and acts, but it has a well-defined quantum physical counterpart: when prehensions are related with perceptions they correspond to preparations in the brain that produce new states in the quantum system, when they are related to volitional acts, they correspond to quantum dispositional states choosing outcomes in the brain--- events---. Therefore, the above requirements can be reduced to two that can be described in physical terms. In fact, if we stick to its physical manifestation, the last two requirements can be replaced by:

ii’) {\em Identity of the sequence of states of the quantum entangled system} Such an identity exists if and only if the sequence of states of the quantum entangled system at t<t' is contained in the sequence of states at t' and any change of state is correlated with an event of preparation or measurement in the brain. 

Requirements i) to iii), and i), ii') are not necessarily equivalent since the causal opening of quantum mechanics admits phenomenal aspects of agents not registered in mere changes of state or quantum leaps but both sets of requirements should be considered as sufficient to ensure personal identity. It is important to note that physical content of these requirements is based on quantum information and not on the matter that is in this state. In quantum mechanics, two systems in the same state are indistinguishable. For example, states can be teleported and information transmitted, and identical people can be based on different material constituents. Furthermore, it is necessary to keep in mind that the definition presented here, although it can be expressed in physical terms, it cannot be expressed in factual terms, since quantum states are not determinable in terms of individual observations, and any attempted observation alters the states without allowing their complete determination. Traditional reductionist definitions that appear in the literature, given purely in terms of psychological or physiological continuity are factually determinable and some of the paradoxes that arise when attempting to define personal identity in terms of them do not apply to the present definition \citep{korfmacher}.

Due to the quantum indeterminism two different beings at $t'>t$ may be causally connected to the same being at $t$.  The second and third conditions are independent in the context of the quantum panprotopsychic vision. Phenomenal continuity ensures that its physical counterpart is an open quantum system whose states are defined by the successive prehensions and therefore its dispositional states are also precisely defined, which in turn ensures that the quantum choices ---the acts--- correspond to a well-defined set of options. Even if one can consider this to be a satisfactory answer to the persistence question that provides a necessary and sufficient set of conditions, what sources of evidence can we have for such a notion of identity?  

The observation of systems capable of being in robust, entangled quantum states in the brain and coupling with the neural system would provide evidence of quantum panprotopsychism and the possibility of free agents. The existence of such states is a necessary condition for the validity of this form of panpsychism as well as the proposed notion of personal identity. On the other hand, this definition implies an extended form of psychological continuity. It includes prehensions and therefore psychological states. But more generally, not only includes the properties of a person as a ``thinking, intelligent being, that has reason and reflection, and can consider itself as itself, the same thinking thing in different times and places” \citep{Locke1689}, II.xxvii.9, It also admits a notion of personal identity that allows the identity of a person at one time with a non-person---in Lockean sense--- at another time. For example, fetuses, infants, or human beings in a persistent vegetative state would meet the above defined conditions \citep{korfmacher}. In fact, fetuses with certain development would have brain activity accompanied by prehensions associated with feelings or sensations. Human beings at the end of their life would certainly have past prehensions and activity in their past and would preserve their neurological organization before brain death.

The first condition ensures classical causal connection of the neurological brain states in two instants of time \citep{shoemaker}. Causal connection does not imply identity between the initial and final state of the brain, but a partial dependence of the state at $t'$ from the state of the brain at $t$. Beyond the psychological continuity postulated by Locke that involves the use of all our mental faculties and is evidenced by memory or another form of record, what source of evidence do we have that ensures that a prehension or act of $S$ at $t$ was an act from $S'$s past? COC relates the informational content of the pre-self-conscious states of the fetus and its brain and its genetic constitution with subsequent psychological developments showing that these requirements lead to a generalized notion of psychological connection. Notice that the physiological connection, when formulated in terms of the informational and organizational content of the brain, is compatible with the expanded form of psychological connection resulting from the aforementioned criteria. However, these criteria, which can be considered reductionist, are not sufficient for a complete description of personal identity. 

As we have seen, a kind of generalization of psychological connection is a consequence of the previous requirements. Arguments have been raised against the correction of psychological identity based on brain fission or teleportation without destruction of the original \citep{parfit}. These arguments do not apply for the kind of psychological identity resulting from our requirements.  For example, if it were possible to teleport a person without destroying the original body, two identical bodies with their corresponding brains and two identical people would be obtained. Since both people would have different locations and different histories after teleportation, one would have two people identical to the original model $S$ ---let us call them $S'$ and $S''$--- who would not be identical to each other $S' \ne S''$ violating the transitivity of identity and therefore showing that psychological continuity is not a good criterion for identity. Something similar would be possible following a procedure called fission in which the brain of $S$ is removed from its body, leaving two (potentially) equipped cerebral hemispheres \citep{korfmacher} that are transplanted into one of two qualitatively identical bodies $S'$ and $S''$, which are assumed to share all psychological characteristics. Both would be candidates to be identical to $S$, but again after fission they will evolve differently having different histories and again leading to a violation of transitivity. Within the context of the quantum panprotopsychism considered here, these characteristics include prehensions and therefore quantum states, which requires the existence of a quantum system of entangled particles where prehensions with their corresponding choices take place. Neither teleportation nor fission can reproduce exact copies of the same system due to the no-cloning theorem of quantum states. In particular, the original quantum state is necessarily destroyed in any physical teleportation process without. With respect to fission, the no-cloning theorem implies that either the entangled system is preserved in only one of the hemispheres or, if in the original brain the entangled system is shared by both hemispheres, it is destroyed during fission and therefore neither $S'$ nor $S''$ would be identical to $S$. Consequently, there is not any violation of transitivity. 
 
Korfmacher has analyzed what he calls the paradox of personal identity. Basically, it does not apply here because psychological continuity is a necessary condition deduced from the aforementioned requirements, but it is not sufficient, a sufficient condition includes physical elements such as quantum states associated with prehensions that cannot be determined by single observations and therefore require non factual evidence. It is not based on facts observable in the third person and therefore his Premise 3 is not satisfied. On the other side his notion of Identity Mysticism is not required and our notion of identity is based on elements that have a physical counterpart. Only a form of psychological continuity that includes prehensions and choices is admissible. Given the impossibility of observing states without destroying them, observation of their corresponding physical counterparts is never sufficient to determine phenomenal contents. This cannot be surprising when phenomenal properties and self-consciousness are constitutive elements of persons.

In a quantum approach to consciousness like the present one, there is no causal closure and different sequences of actions would be possible. Rigorous causal connection is only probabilistic and different agents at time $t'$ could be psychologically causally connected with a subject at $t<t'$.  If it is admitted that there is freedom to act otherwise, the above defined classical organizational connection is not sufficient to ensure personal identity between different agents. It should be noted that psychological continuity is not necessary to satisfy the condition of the identity of agents. Agents do not require psychological continuity; even death and resurrection later could be consistent with IA or PC without meeting any psychological continuity condition. However, it must be emphasized that actions and choices can occur only in a physical system such as the person's brain. Disembodied choices do not seem to make any sense, and one cannot speak of personal identity beyond the duration of one's life. Perceptions and their corresponding prehensions are required. But as Cassirer observes \citep{cassirer2} p. 121 “Perception is the only thing that discloses reality… Perception gives us the only (immediate) insight concerning reality , something which can never be obtained from conceptual, logical means” Without the information provided by our perceptions, the acts we refer to would have no meaning.  Actions become relevant when freedom is possible and reasons and motivations play a role in decision making. In choices and actions, we include not only the pursuit of ends for their own sake but also intentions, that is, choosing means to achieve ends and any choice of a line of action or a particular way of thinking. \citep{kenny}. Significant acts resulting from free choices are compatible with probabilistic descriptions such as that provided by quantum mechanics. The frequency with which letters appear in words in the English language is known, with 'e' being the letter that appears most frequently. Different texts with different meanings can be transmitted with letters that appear with equal frequency in both. Thus probabilistic predictions are compatible with freedom. 

\section{Conclusions}

Non-human animals operate based on biologically predetermined goals they cannot change, whereas human beings can assess and alter their goals. This indicates a strong connection between self-consciousness, self-determination, and free will in humans. This idea is further supported by the quantum panprotopsychist approach where the role of quantum choices in determining outcomes results from the probabilistic nature of quantum mechanics. The enduring aspect of identity is not found in our moment-to-moment selves but in the series of prehensions and free actions that address our objectives, existential concerns, and moral responsibilities.
The process of reflection leading to a free act of the self, which aligns with the form of quantum panpsychism proposed here, involves three levels: the information stored in our brain, the prehensions related to the topic requiring the decision and the agent responsible of the choice among the different possibilities regarding the topic in question. The above-mentioned levels involved in the process of reflection determine someone’s personal identity at a certain time and answers the so-called characterization question.

The three levels that intervene in the process of reflection followed by a free action of the "I” are involved in the self’s manifestations, but actions are the main cause of the development of our brain capacities and of its present activities, and therefore the defining elements of the self. Thus, we have proposed three requirements for the persistence of the personal identity: organizational connection of the brain,identity of subjects of experience and identity of agents. The first condition ensures the causal connection between brain states at two different instants of time \citep{shoemaker}. This causal connection does not imply that the initial and final states of the brain are identical, but rather that the state at $t'$ partially depends on the state at $t$, as they are two stages of the same 'object'. The second requirement ensures the phenomenal continuity of the subjects of experience and the third is an independent criterion arising from the probabilistic nature of the quantum panprotopsychic view.

The paradox presented by Korfmacher \citep{korfmacher}, applies to most solutions to the persistence problem through a series of thought experiments demonstrating the challenges confronted by the physicalist definitions of physiological or psychological continuity. He assumes, consistently with a classical physics-based ontology, the causal closure of the physical world and, consequently, its incompatibility with any non-reductionist proposal such as the Simple View. However, this paradox does not hold in the quantum panprotopsychism framework because: a) fission and teleportation with preservation of the original system are impossible due to the no-cloning theorem, thus ensuring well-defined psychological continuity; and b) in this quantum framework, psychological continuity including the succession of prehensions and identity of agents are compatible criteria that, when combined, provide necessary and sufficient conditions for the persistence of personal identity.

The activity of the "I", which is purely of phenomenal origin and the source of our evaluations and choices, is the fundamental origin of all cultural expression. This activity reveals an internal law that is entirely analogous to the law governing nature but operates at a higher level of necessity.  \citep{cassirer2}.The panpsychist vision presented here offers a renewed perspective of the world focused on its phenomenal development and the constitution of agents endowed with self-consciousness that give rise to a new social and cultural dimension based on people's inherent capacity for conceptual elaboration and symbolic representation. Understanding the emergence of these artistic, moral, and cognitive capacities in a panpsychist context will be the subject of subsequent work. Furthermore, we are also interested in exploring possible test scenarios and empirical corroborations of the existence of quantum processes, as predicted by this approach.


\end{document}